\documentclass[12pt, fleqn] {article}
\begin{document}

\title{\bf Relativistic Calculations of Induced Polarization
           in $^{12}C\left(e, e^\prime \vec{p}\right)$ Reactions\thanks{
           This work supported in part by the Natural Sciences
           and Engineering Research Council of Canada} }

\author{ {\bf J.I. Johansson and H.S. Sherif} \\
              Department of Physics, University of Alberta \\
              Edmonton,  Alberta, Canada T6G 2J1}

\date{\today}

\maketitle

\begin{abstract}

Relativistic calculations of the induced proton polarization in
quasifree electron scattering on $^{12}C$ are presented.
Good agreement with the experimental data of Woo {\em et al.} is obtained.
The relativistic calculations yield a somewhat better description of the
data than the non-relativistic ones.
Differences between the two approaches are more pronounced at larger
missing momenta suggesting further experimental work in this region.

\end{abstract}


\newpage


A measurement of induced proton polarization in the
$^{12}C\left(e, e^{\prime} \vec{p}\right)$ reaction has been reported recently by
Woo {\em et al.} \cite{Wo98}.
The data explore the low missing momentum region from
0 to 250 MeV/c in constant q-$\omega$ kinematics.
The data were compared to non-relativistic DWIA calculations using the
effective momentum approximation.
Momentum distributions used in these calculations were obtained by fitting the
$^{12}C\left(e, e^{\prime}p\right)$ data of van der Steenhoven {\em et al.}
\cite{St88}.
The final state interactions of the outgoing proton were included using
non-relativistic optical potentials.
Two such potentials were compared, one obtained by a
reduction of the relativistic potentials of Cooper {\em et al.} \cite{COPE},
and the other based on an empirical effective interaction (EEI).
Both models provide reasonable agreement with the data, with a slight
preference for the EEI model when a proton is removed from the $1p_{3/2}$
shell.
The results from both non-relativistic models pass through the lowest missing
momentum point for the knockout of a $1p_{3/2}$ proton.
The change in polarization with increasing missing momentum is reproduced,
but the calculations tend to fall below the data at higher missing momenta.
The calculations predict that the polarization should rapidly become negative
as the missing momentum is increased beyond 220 MeV/c.

For the set of data attributed to knockout of a $1s_{1/2}$ proton,
the non-relativistic calculations follow the trend of the data for small missing
momenta, but become large and negative for missing momenta
beyond 150 MeV/c.
The data, on the other hand, seem to indicate a polarization becoming
positive at these larger values of missing momenta.

There also exist relativistic calculations for the same reaction, which
have primarily considered the cross section, or equivalently, the
single--particle momentum distribution.
These calculations have mainly been reported by
Udias {\em et al.} \cite{Ud93,Ud95}, Jin and Onley \cite{JO94}
and Hedayatipoor {\em et al.} \cite{HJS95}.

Relativistic calculations of proton polarization for the reaction
$^{16}O\left(e, e^{\prime} \vec{p}\right)$,
have been reported previously by Johansson and Sherif \cite{JS97}.

In the present paper we compare our full relativistic calculations
to the new data presented in reference \cite{Wo98}.
We also point out differences between the results of the current model
and those discussed in \cite{Wo98}.

In the following text we outline the relativistic calculations for the
quasifree electron scattering reaction, and discuss how the
proton polarization is calculated.
The results of our relativistic calculations are then presented along with
discussion of the data and calculations of Woo {\em et al.}.


The relativistic calculations of the amplitude, in the one photon
exchange model for the $\left(e, e^{\prime}p\right)$ process,
are discussed in references \cite{HJS95,JS96}.
The main results are given briefly here in the notation of Johansson
and Sherif \cite{JS96}.
(Note that we use the de Forest prescription, cc2,
as discussed in ref. \cite{HJS95}.)
We do not include Coulomb distortion in the leptonic part of the
amplitude as this is not expected to be important for the light nucleus
considered here \cite{Mc90,GP87}.

The relativistic expression for the differential cross
section leading to a specific final state of the residual
nucleus can be written as
\begin{eqnarray}
 \frac{d^3\sigma}{d\Omega_{p} d\Omega_f dE_f}  &=&
   \frac{2} {\left( 2 \pi \right)^{3}}  \frac{\alpha^{2}} {\hbar c}
   { \left[ \frac{ \left( m_e c^{2} \right)^{2}
                   M c^{2} \;
                   {\left| \mbox{\boldmath{$p$}}_{p} \right|} c}
                 { \left[ \left( q c \right)^{2} \right]^{2}}
             \frac{ {\left| \mbox{\boldmath{$p$}}_{f} \right|} c}
                  { E_i }
            \right] }
    \frac{ c } {v_{rel} } \frac{1}{R}
  \nonumber
   \\  & & \times { \frac{ {\cal S}_{J_i J_f} (J_B) }{ 2J_B + 1 } }
    \sum_{ \mu M_B \nu_{f} \nu_{i} }
    { \left| e_{\nu_{f} \nu_{i}}^{\beta}
             N_{\beta}^{\mu M_B} \right| }^2 ,
  \label{cross}
\end{eqnarray}
where $\nu_i$ and $\nu_f$ are the spin projections of the incoming
and outgoing electrons respectively, while $M_B$ and $\mu$ are the
spin projections of the bound and continuum protons.
The 4-momenta of the initial and final electrons are
$p_{i}$ and $p_{f}$ respectively, while the final proton
4-momentum is $p_{p}$. The 4-momentum of the exchanged photon
is $q$ and is calculated as the difference between the
initial and final electron 4-momenta $q = p_{i} - p_{f}$.
The 4-momentum of the recoil nucleus is $p_{R}$, and the initial
4-momentum of the struck proton, denoted $p_{m}$,
is often referred to as the missing momentum.
The recoil factor $R$ is given in any frame by \cite {FM84}
\begin{eqnarray}
   R = 1 - \frac{ E_{p} } { E_{R} }
           \frac{1} {\left| \mbox{\boldmath{$p$}}_{p} \right|^2}
           \mbox{\boldmath{$p$}}_{p} \cdot
           \mbox{\boldmath{$p$}}_{R}     .
   \label{recoil_fact}
\end{eqnarray}
The matrix element, $N_{\beta}^{\mu M_B}$, which involves nuclear wave
functions, can be written in the form
\begin{eqnarray}
  N_{\beta}^{\mu M_B} = \int d^3x \;
        \Psi_{\mu}^\dagger \left( p_{p}, \mbox{\boldmath{$x$}} \right)
        \Gamma_{\beta}
        \Psi_{J_{B}, M_{B}} \left( \mbox{\boldmath{$x$}} \right)
        \exp \left( i \mbox{\boldmath{$q$}} \cdot
                      \mbox{\boldmath{$x$}} \right)     ,
   \label{nuc_mat}
\end{eqnarray}
where the wave functions of the continuum and bound nucleons,
denoted $\Psi_{\mu}$ and $\Psi_{J_{B}, M_{B}}$ respectively, are
solutions of the Dirac equation containing appropriate potentials
\cite{LS88}.
The $4 \times 4$ matrix $\Gamma_{\beta}$, operating
on the nucleon spinors, and the four-vector which comes from the
electron vertex, $e_{\nu_{f} \nu_{i}}^{\beta}$ in Eq.
(\ref{cross}), are given in detail in Eqs. (2.8) and (2.9) of
reference \cite{HJS95}.

We define a matrix element which is a function of the spin projections
of the initial and final particles by the relation
\begin{eqnarray}
  T_{\nu_{f} \nu_{i}}^{\mu M_B}
     = e_{\nu_{f} \nu_{i}}^{\beta} N_{\beta}^{\mu M_B}   ,
   \label{T_mat}
\end{eqnarray}
where summation over $\beta$ is implied.
The polarization of the proton along the y-axis can then be written as
\begin{eqnarray}
  P_{y} = -2 \frac{ \mbox{Im}
                    \sum_{ M_B \nu_{f} \nu_{i} }
                       T_{\nu_{f} \nu_{i}}^{1/2 M_B}
               \left[ T_{\nu_{f} \nu_{i}}^{-1/2 M_B}
                     \right]^{\ast}
                  }
                  { \sum_{ \mu M_B \nu_{f} \nu_{i} }
               \left| T_{\nu_{f} \nu_{i}}^{\mu M_B}
                     \right|^{2}
          }   .
   \label{prot_pol}
\end{eqnarray}
We utilize kinematics in which the incident electron momentum defines the
z-axis, and the final electron momentum lies in the x-z plane with $\phi_{e}$
= 0$^\circ$, while the final proton momentum has $\phi_{p}$ = 180$^\circ$.
This means that the polarization which we calculate is the negative of the
polarization reported by Woo {\em et al.}.
We multiply our calculated polarization $P_{y}$ by the factor ($-1$) to
conform with the sign convention of Woo {\em et al.}


The present relativistic calculations use bound state wave functions
generated using the Hartree potentials of Blunden and Iqbal \cite{BI87}.
We also use a bound state wave function generated using phenomenological
Woods-Saxon potentials in order to test the sensitivity of the results to
changes in the description of the bound state.
The proton optical potentials are taken from Cooper {\em et al.} \cite{COPE}.
There are several sets of optical potentials available, some of
which are energy dependent (E--dep) and constructed from a fit
to data for a specific nucleus, such as $^{12}C$, $^{16}O$
and $^{40}Ca$, in the proton kinetic energy
range of $\sim$ 25 $MeV$ to 1 $GeV$.
Other potentials are parameterized in terms of target mass as
well as proton energy (E+A--dep) and can be used to generate
potentials for which no proton elastic scattering data exist.
We shall perform calculations using both types of potentials.

Our results are compared in Fig. 1 with the non-relativistic EEI
calculations of ref. \cite{Wo98}.
The values of $q$ and $\omega$ used in the non-relativistic calculations
were provided to us by Kelly \cite{JK98}.
Our values of $q$ and $\omega$ are consistent with those used in the
non-relativistic calculations if we ignore the electron mass.
For the $1p_{3/2}$ case, Fig. 1(a), the relativistic calculations provide
a slight improvement over the non-relativistic results.
This is true for both types of optical potentials used in the relativistic
calculations.
We note that the differences between relativistic and non-relativistic
calculations are accentuated at higher missing momenta.
There are also differences at very small $p_{m}$, with the
relativistic model providing a 10\% smaller polarization at missing
momenta around 20 MeV/c.

The calculations for $1s_{1/2}$ proton knockout are shown in Fig. 1(b).
It is more evident in this case that the relativistic calculations
provide better agreement with the data, particularly at large missing
momenta.
(They are also better than the non-relativistic calculations using the
EDAIC potential reported in Ref. \cite{Wo98}, but not shown here).
The most prominent feature here is that the non-relativistic calculations
with the EEI potential produce a large negative polarization in the
region $p_{m}$ = 250 - 300 MeV/c.
The EDAIC potential produces a shallower minimum in the same region.
By contrast the relativistic calculations indicate that the minimum
would be at missing momenta larger than 300 MeV/c.
Because of the large size of the error bars for the data points at
the largest missing momenta, the behaviour of the polarization is not
well constrained in this region, but the relativistic calculations
seem to be following the trend of the data somewhat better than the
results of the non-relativistic model.

Sensitivity to changes in the binding potentials has been
examined by performing calculations using both Dirac Hartree and
Woods-Saxon binding potentials.
The Hartree potentials result in a binding energy that is slightly
smaller than the experimental value.
The Woods-Saxon potentials reproduce the experimental
binding energy and also provide an $rms$ radius for the bound state
that is within one percent of that found from the Hartree potentials.
We find little difference between the two binding potentials for the
region of missing momenta considered in the present study.
This occurs because the momentum-space wave functions for these bound
states are very similar in the low momentum region explored in these
kinematics.

We stress that the issue here is not one of final state interactions,
since we are using the same potentials used by Woo {\em at al.}.
The issue is the difference between results arising in the relativistic
and non-relativistic treatments of these reactions.
We have thus shown that the relativistic calculations appear capable of
achieving better agreement with the nucleon polarization data than the
non-relativistic ones.
This is consistent with observations made earlier in polarized nucleon
scattering experiments.
We also note a hint of large differences between the two calculations
at large missing momenta.
This suggests the advisability of pushing the measurements further into
the high missing momentum region.
It is expected that such measurements would strongly test both models and
also clarify the role, if any, of two body currents in these reactions.

\section*{Acknowledgements}

JIJ would like to thank James Kelly for kindly providing his EEI
calculations from reference \cite{Wo98}.

\newpage

\section* {Figure Captions}

\noindent FIG. 1. Polarization of the knocked-out proton in the
$^{12}C\left(e, e^{\prime} \vec{p}\right)^{11}B$ reaction.
The energy of the incident electron is 579 $MeV$,
with constant q-$\omega$ kinematics.
The Hartree bound state wave functions are from \cite{BI87}
while the proton optical potentials are from reference \cite{COPE}.
(a) Knockout of a $1p_{3/2}$ proton.
(b) Knockout of a $1s_{1/2}$ proton.
Solid curves  --- Hartree binding potential and
                  E-dep optical potential for $^{12}C$.
Dashed curves --- Woods-Saxon binding potential and
                  E-dep optical potential for $^{12}C$.
Dotted curves --- Hartree binding potential and
                  E+A-dep optical potential, fit 1.
Dot-dashed curves --- EEI calculations from \cite{Wo98}.
The data are from reference \cite{Wo98}.
Closed circles denote missing energy in the range $28<E_{m}<39$ MeV, and
open circles denote missing energy in the range $39<E_{m}<50$ MeV.

\begin{figure}
\begin{picture}(1100,400)(0,0)
\includegraphics{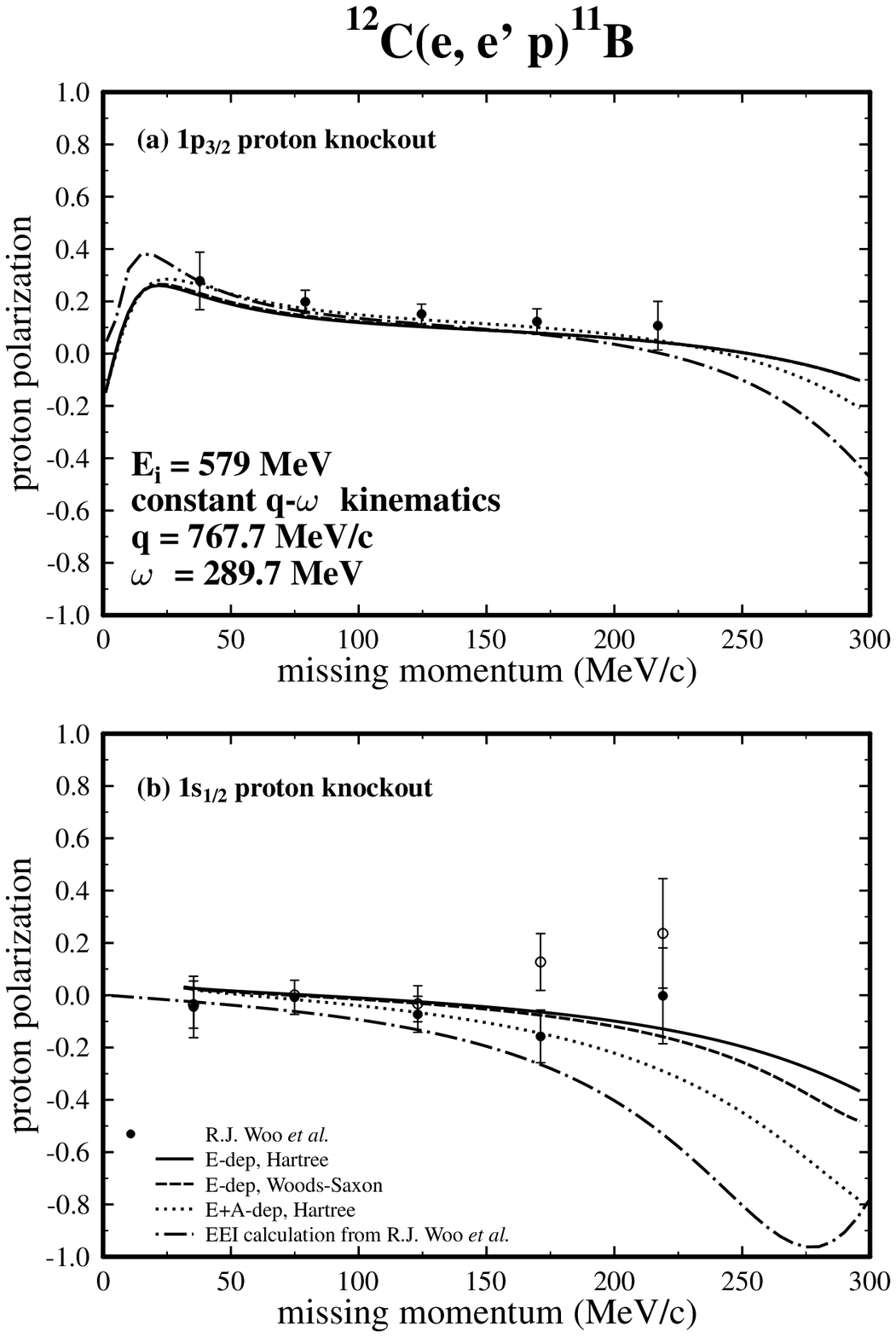}
\end{picture}
\end{figure}

\end{document}